\documentclass[fleqn, twocolumn, 11pt]{wlscirep}
\usepackage[utf8]{inputenc}
\usepackage[T1]{fontenc}
\title{Pink-Beam Dark Field X-ray Microscopy: Expanding 3D/4D Imaging for Complex and Deformed Microstructures}
\usepackage{gensymb}
\usepackage{multirow}
\usepackage{ragged2e}
\usepackage{xcolor}
\definecolor{RubineRed}{rgb}{0.82, 0.0, 0.34}
\def\tablename{Table}%
\usepackage{float}
\usepackage{amsmath} 
\usepackage{epstopdf}           %
\newcommand{\un}[1]{\ensuremath{\,\mathrm{#1}}}

\usepackage{lmodern}
\usepackage{siunitx}

\newcommand{\pDFXM}[0]{{\textcolor{RubineRed}p}DFXM}

\usepackage{lineno}
\usepackage[figuresright]{rotating}
\usepackage{color,soul}

\usepackage{hyperref}
\hypersetup{
    colorlinks=true,
    linkcolor=blue,
    filecolor=magenta,      
    urlcolor=cyan,
    pdftitle={Overleaf Example},
    pdfpagemode=FullScreen,
    }
    
\usepackage{subfigure}

\author[1,*]{Can Yildirim}
\author[1]{Aditya Shukla}
\author[2]{Yubin Zhang}
\author[3]{Nikolas Mavrikakis}
\author[1]{Louis Lesage}
\author[1]{Virginia Sanna}
\author[1]{Marilyn Sarkis}
\author[1]{Yaozhu Li}
\author[4]{Michela La Bella}
\author[1]{Carsten Detlefs}
\author[4]{Henning Friis Poulsen}

\affil[1]{European Synchrotron Radiation Facility, 71 Avenue des Martyrs, CS40220, 38043 Grenoble Cedex 9, France.}
\affil[2]{Department of Civil and Mechanical Engineering, Technical University of Denmark, 2800 Kgs. Lyngby, Denmark}
\affil[3]{OCAS, Pres. J.F. Kennedylaan 3, BE-9060, Zelzate, Belgium}
\affil[4]{Department of Physics, Technical University of Denmark, 2800 Kgs. Lyngby, Denmark }

\affil[*]{can.yildirim@esrf.fr}

\begin{abstract}
Dark Field X-ray Microscopy (DFXM) has advanced 3D non-destructive, high-resolution imaging of strain and orientation in crystalline materials, enabling the study of embedded structures in bulk. However, the photon-intensive nature of monochromatic DFXM limits its applicability to highly deformed or weakly crystalline structures and constrains time-resolved studies in industrially relevant materials.
We present pink-beam DFXM (\pDFXM) at the ID03 beamline of ESRF, achieving a 27-fold increase in diffracted intensity while maintaining 100 nm spatial resolution. We validate \pDFXM{} by imaging a partially recrystallized aluminum grain, confirming sufficient angular resolution for microstructure mapping. The increased flux significantly enhances the diffracted signal, enabling the resolution of subgrain structures. Additionally, we image a highly deformed ferritic iron grain, previously inaccessible in monochromatic mode without focusing optics.
Beyond static imaging, \pDFXM{} enables real-time tracking of grain growth during annealing, achieving hundred-millisecond temporal resolution. By combining high photon flux with non-destructive, high-resolution 3D mapping, \pDFXM{} expands diffraction-contrast imaging to poorly diffracting crystals, unlocking new opportunities for studying grain growth, fatigue, and corrosion in bulk materials.

\end{abstract}

\begin{document}
\flushbottom
\maketitle
\section*{Introduction}
In crystalline materials, such as metals,  geological materials, semiconductors and biominerals, material properties are closely linked to internal defect structures like dislocations and to the preferred orientation, or crystallographic texture, of grains. These features are critical because they govern how materials respond to stress, wear, corrosion and other external forces. To fully understand the mechanisms that drive performance, it is essential to examine these characteristics in three dimensions (3D), as real-world materials exhibit complex, multi-scale bulk heterogeneities that profoundly impact their behavior.

Traditional microscopy methods like Transmission Electron Microscopy (TEM) and Electron Backscatter Diffraction (EBSD) have transformed our understanding of microstructural features \cite{pantleon2008resolving, moussa2015, Echlin2012, Huang1997}, but they are limited to surface observations or thin slices, unable to access the full volume of bulk samples in a non-destructive way. This restriction has led to the development of advanced 3D X-ray techniques, including Three-Dimensional X-ray Diffraction (3DXRD) \cite{poulsen2004three,Schmidt229, hefferan2012}, Diffraction Contrast Tomography (DCT) \cite{king2008observations, sun2018}, and Differential Aperture X-ray Microscopy (DAXM) \cite{xu2017direct, knipschildt2025}. Using the penetration powers of hard X-rays, these methods enable non-destructive, volumetric imaging, providing crucial multi-grain mapping and insight into grain orientation within bulk materials. However, each faces certain limitations. Standard 3DXRD and DCT are constrained by diffraction peak overlap at high plastic strain levels, and they lack the spatial resolution required to resolve sub-grain deformation features. Emerging adaptations like scanning 3DXRD \cite{henningsson2024, shukla2024grain} and texture tomography \cite{frewein2024texture, carlsen2024x} overcome the peak overlap problem and can improve spatial resolution but rely on time-consuming raster scanning, limiting their ability to monitor dynamic events. A similar limitation exists for DAXM, where the need for raster scanning constrains temporal resolution in tracking fast microstructural changes. Higher spatial resolution methods, such as Bragg coherent diffraction imaging \cite{richard2022bragg} and Bragg ptychography \cite{hruszkewycz2017high}, can map single crystals and nanoparticles with a resolution of a few tens of nanometers. However, they are limited to low-defect-density crystals in non-extensive samples, such as isolated particles or thin films, where reconstructing the diffracted signal to generate 3D maps becomes increasingly challenging for complex materials with high defect densities.

Dark Field X-ray Microscopy (DFXM) emerged as a new tool for extensive full-field imaging that complements grain mapping methods like 3DXRD and DCT. While multi-grain methods provide a statistical overview of bulk microstructures, DFXM is well-suited for resolving intragranular 3D information with superior angular sensitivity and spatial resolution (on the order of 100 nm)\cite{Simons2015,yildirim2020probing}. Combining hard X-ray diffraction with a full-field imaging configuration at photon energies of 15--35keV, DFXM enables non-destructive visualization of strain and orientation variations within individual grains, capturing deformation processes across different length scales. For a detailed description of the technique, refer to Refs.~\citeonline{Poulsen2017,isern2024}. Recent studies using DFXM have demonstrated the 3D mapping of recrystallized grains, deformed microstructures, single crystals, and domain structures in ferroelectric materials, capturing details of subgrain structures, dislocation patterns, phase transformations, and domain switching\cite{simons2018,Dresselhaus2021, extensive, Mavrikakis2019, yildirim2023exploring, Ahl2017, Bucsek2019}.

However, DFXM also presents distinct limitations when compared to 3DXRD, DCT, and DAXM. First, its single-grain imaging approach limits information about neighboring grains, making it challenging to place observations within the broader microstructural context. One possible solution is the integration of 3DXRD or DCT with DFXM, enabling simultaneous multi-grain context and high-resolution intragranular imaging. This approach has been demonstrated in a few pioneering experiments \cite{gustafson2020,gustafson2023, chen2023high}, but it remains technically demanding and is beyond the scope of this paper.

Second, and most critically, DFXM is a notoriously photon-demanding technique, requiring high-intensity synchrotron radiation to resolve weakly diffracting features. This severely limits its use in highly textured materials, such as deformed metals, shock-deformed geological samples, and biominerals, where low signal-to-noise ratios make it challenging to map grains with high dislocation densities and/or complex hierarchical structures. While previous studies have demonstrated the feasibility of mapping deformed microstructures \cite{yildirim20224d}, these experiments remain highly challenging due to long exposure times and low diffraction efficiency. 
For grains subjected to high to severe plastic deformation\cite{Valiev2000} (von Mises strain > 0.5) , larger orientation spreads necessitate wider angular scan ranges, significantly extending data collection times. As a result, mapping the full 3D orientation of a deformed crystal can take tens of hours, making systematic studies impractically slow.

Addressing the diffracted intensity limitations of monochromatic DFXM required a significant technological advancement. Here, we present the first application of pink-beam DFXM (\pDFXM), enabled by the new ID03 beamline at the European Synchrotron Radiation Facility (ESRF) as part of the Extremely Brilliant Source (EBS) upgrade \cite{isern2024, cloetens2025esrf}. Unlike conventional monochromatic beams (\(\Delta E/E \sim 10^{-4}\)), the pink beam has a broader energy bandwidth (\(\Delta E/E \sim 10^{-2}\)), leading to a \(\sim 100\)-fold increase in incoming X-ray flux. This increase enables rapid, high-resolution 3D mapping of weakly diffracting, highly deformed, and textured materials. While pink-beam illumination has been used in bright-field X-ray microscopy\cite{falch2016correcting}, its application in diffraction mode with compound reflective lenses has not been explored until now. 

\begin{figure*}[t]
    \centering
    \includegraphics[width=0.78\textwidth]{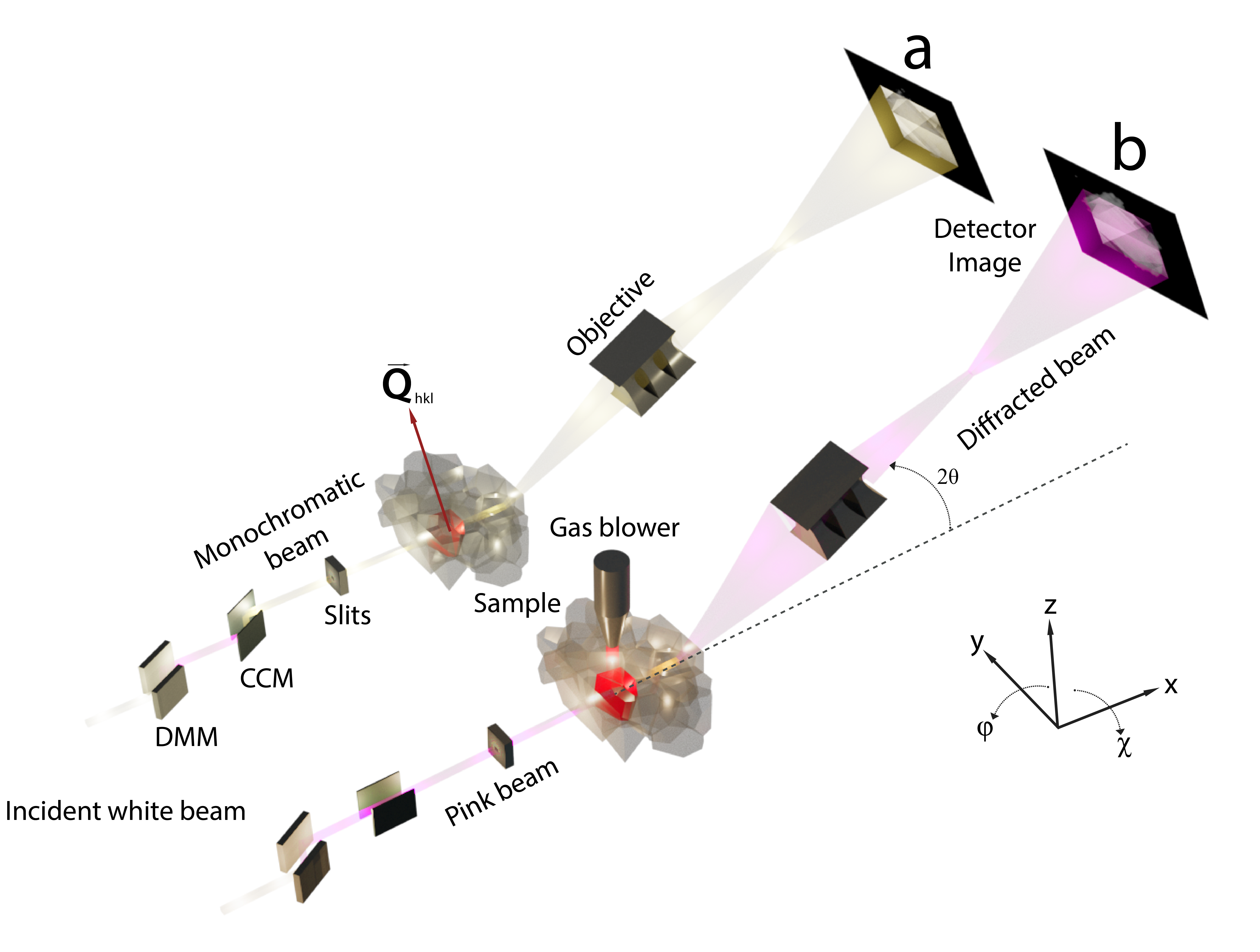} 
    \caption{\small \justifying \textbf{Schematics of monochromatic and pink-beam dark-field X-ray microscopy (\pDFXM) setups.} 
    (a) Monochromatic DFXM configuration, where a channel-cut monochromator (CCM) selects a highly monochromatic beam with an energy bandwidth of $\Delta E / E = 1.4 \times 10^{-4}$ The diffracting grain of interest (GOI) is indicated in red. The objective lenses magnify the diffracted signal from the GOI. DFXM orientation maps are obtained by scanning two tilt angles (\(\phi\) and \(\chi\)) at a fixed scattering angle \(2\theta\), revealing spatial variations in lattice orientation around the \( Q_{hkl} \) scattering vector.  
    (b) \pDFXM\ configuration, where a double multilayer monochromator (DMM) selects a broader bandwidth bypassing the CCM. The increased bandwidth leads to a larger spread of the diffracted beam in the \(2\theta\) direction. A gas blower (shown in b) was used to perform in-situ annealing of a forged aluminum sample, leveraging the \textit{in-situ} imaging capabilities of \pDFXM\ to enable real-time monitoring of grain growth.
    The transition between monochromatic and \pDFXM\ configurations is achieved in \textit{less than a minute} by shifting a few motors with an 8 mm offset in the lab $y$ direction, allowing seamless switching between the two modes.}
    \label{fig:dfxm_schematics}
\end{figure*}

\begin{figure*}[t]
    \centering
    \includegraphics[width=1\textwidth]{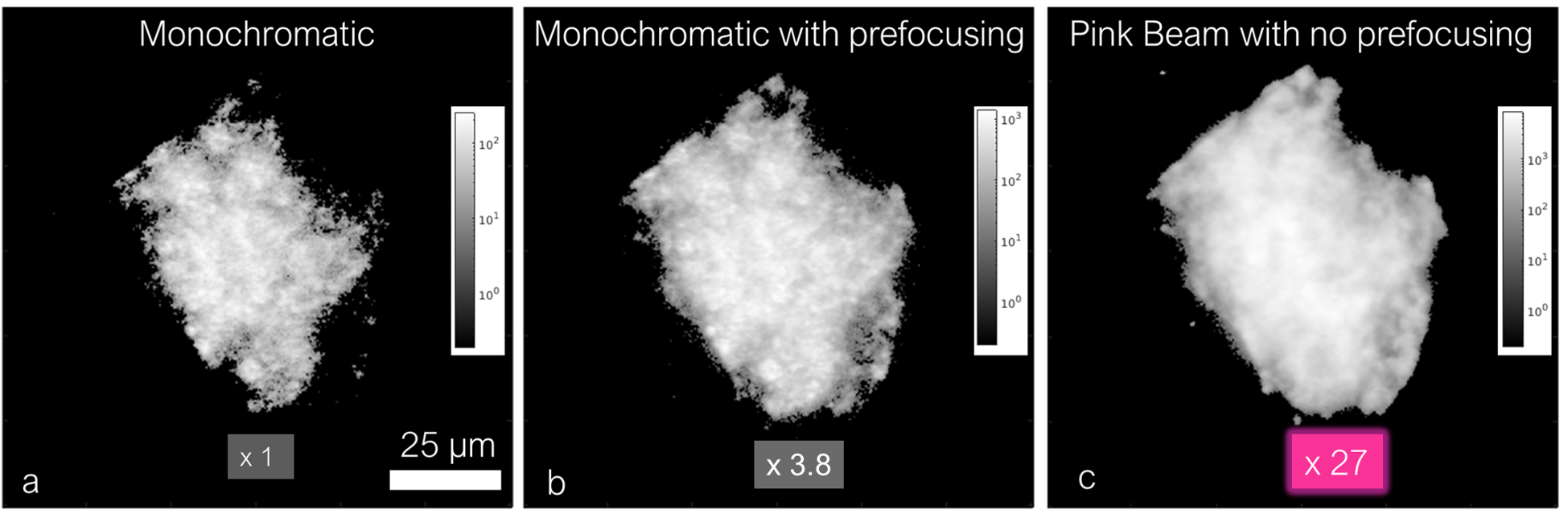} 
    \caption{\small \justifying \textbf{DFXM intensity comparison using different beam configurations.} 
    Raw intensity projections of an aluminum grain in a partially recrystallized sample, acquired using (a) a monochromatic beam, (b) a monochromatic beam with prefocusing via transfocator (2D compound refractive lenses, CRLs) placed between CCM and slits shown in Fig.~\ref{fig:dfxm_schematics}, and (c) a pink beam without prefocusing. The images are presented on a logarithmic intensity scale, with relative intensities normalized to the monochromatic beam. The 27-fold intensity increase in (c) demonstrates the potential of \pDFXM\ for improved signal collection and reduced acquisition times in 3D microstructure characterization using DFXM. }
    \label{fig:int_comp}
\end{figure*}

In this work, we systematically compare \pDFXM{} and monochromatic DFXM, evaluating diffracted intensity, orientation mapping, and angular resolution. \pDFXM{} achieves at least a 27-fold intensity increase, resolving recrystallized and highly deformed microstructures. We also demonstrate the first time-resolved \pDFXM{} study of grain growth during recrystallization, capturing real-time boundary migration and intragranular orientation evolution. While \pDFXM{} is not well suited for elastic strain measurements due to increased bandwidth, its rapid switching with monochromatic mode enhances experimental flexibility, ensuring optimal imaging for diverse applications. These results extend DFXM beyond traditional applications, unlocking new opportunities for studying crystalline materials across fields from metallurgy to biominerals.

\section*{Results}
\subsection*{\textit{Experimental Setup}}
In the monochromatic DFXM configuration\cite{isern2024} (Figure~\ref{fig:dfxm_schematics}a), the undulator beam is first reflected by a double multilayer monochromator (DMM) before being refined by a channel-cut Si(111) monochromator (CCM), selecting a highly monochromatic beam with a bandwidth of \(\Delta E / E = 1.4 \times 10^{-4}\). DFXM experiments typically use photon energies between 15-–35 keV. The diffracted beam from a grain of interest (GOI) passes through an objective lens (Be compound refractive lens, CRL), forming a magnified real-space image on a high-resolution detector for structural, orientation, and strain mapping. Due to the objective's numerical aperture (NA), only a fraction of the diffracted beam is captured, acting as a filter in angular space. More details on the technique can be found in Refs.~\citeonline{Poulsen2017,isern2024}.  

In the \pDFXM\ configuration (Figure~\ref{fig:dfxm_schematics}b), the CCM is removed from the beam path, allowing the pink beam to reach the sample with a \(-8\) mm offset in the \(y\) direction\cite{isern2024}. 
Unlike the narrow-bandwidth monochromatic beam,  the pink beam has a significantly broader spectrum, with \(\Delta E / E = 12\%\) FWHM in the 12–24 keV range, decreasing to 2.5\% at 30-–60 keV \cite{isern2024}. This broader bandwidth increases photon flux, reducing exposure times significantly and improving the signal-to-noise ratio, particularly for weakly diffracting or highly deformed structures. However, the increased bandwidth degrades the longitudinal $q$-resolution ($2\theta$), thus limiting elastic strain measurements\cite{Poulsen2017}. 
Furthermore, the increased photon flux on the sample may increase radiation damage and beam heating \cite{lawrence2021}. 
Despite this, the higher intensity of the pink beam enables imaging of microstructures that are otherwise difficult to resolve with monochromatic DFXM.  

For demonstration purposes in this manuscript, all prealignment was performed using the monochromatic beam. Once the desired diffraction condition was established, the system was switched to the \pDFXM{} mode by translating the CCM out of the beam path. This switch between modes is highly reliable, as it only requires shifting a few motors by a predefined offset. 

In the following sections, we present results to demonstrate the imaging capabilities of pink beam mode and validate them with monochromatic counterparts. All presented results use box-beam illumination, which shows a projection of a diffracted grain on the detector.  Orientation maps are acquired using mosaicity scans, where the sample is rotated in both the $\phi$ (rocking) and $\chi$ (rolling) directions to map the distribution of intensity in each voxel. A 2D Gaussian fit is applied to extract the Center-of-Mass (COM) positions, which are used to generate local pole figure representations of orientation variations around a diffraction vector\cite{Garriga2023}.  Additionally, rocking curves are obtained by scanning only in the $\phi$ direction while keeping $\chi$ constant. These curves are analyzed using 1D Gaussian fits to determine average $\phi$ orientation (peak position maps), which provide insights into local lattice rotations and strain variations.  More details on the data analysis are provided the methods section.

\subsection*{\textit{DFXM Intensity Comparison Between Monochromatic and Pink Beam}}
In Figure~\ref{fig:int_comp}, we compare raw DFXM images after background removal and quantify the diffracted intensities collected through the objective lens at 1 second exposure for three different illumination modes. The sample is a recrystallized aluminum grain ($\approx$40 $\mu$m in size) imaged under (i) monochromatic parallel beam illumination, (ii) monochromatic beam prefocused using diamond compound refractive lenses (CRLs), and (iii) pink-beam illumination with no prefocusing. All images correspond to the same grain in the 111 reflection, recorded at a fixed \(2\theta\) and sample tilts ($\phi$ and $\chi$). 

Panel a shows the image acquired with a monochromatic parallel beam (no prefocusing), where significant portions of the grain remain invisible due to insufficient diffracted intensity. This image serves as the baseline for comparison. Panel b presents the image obtained using prefocusing with diamond CRLs in the monochromatic beam configuration, which enhances the mean diffracted intensity by a factor of 3.8. Previously undetectable regions of the grain become visible, and the overall image contrast improves, highlighting the benefits of beam conditioning in standard monochromatic DFXM. Panel c, however, shows the \pDFXM{} image (without any prefocusing), revealing a remarkable 27$\times$ increase in the mean diffracted intensity. Other regions of the grain are now visible at the same fixed orientation and 2$\theta$, with a significantly improved intensity distribution. The estimated photon flux on the sample for each case is given in the methods section. 
This enhancement arises not only from the higher incident flux but also from the increased bandwidth of the pink-beam illumination, which coarsens the angular resolution just enough to integrate over a small angular range in a single image, and also suppress dynamical diffraction fringes. This effect is especially useful for imaging near-perfect crystals, such as recrystallized grains and thick single crystals, where dynamical diffraction artifacts can interfere with integrated intensity maps. It works similarly to the suppression of dynamical diffraction fringes in TEM imaging \cite{wang1995dynamical}.This makes it a powerful tool for high-throughput microstructural studies. See the SI for a detailed comparison of DFXM angular resolution under monochromatic and pink-beam conditions, and reduction of the dynamical diffraction fringes on a fully recrystallized iron sample.

\begin{figure}[t]
    \centering
    \includegraphics[width=0.5\textwidth]{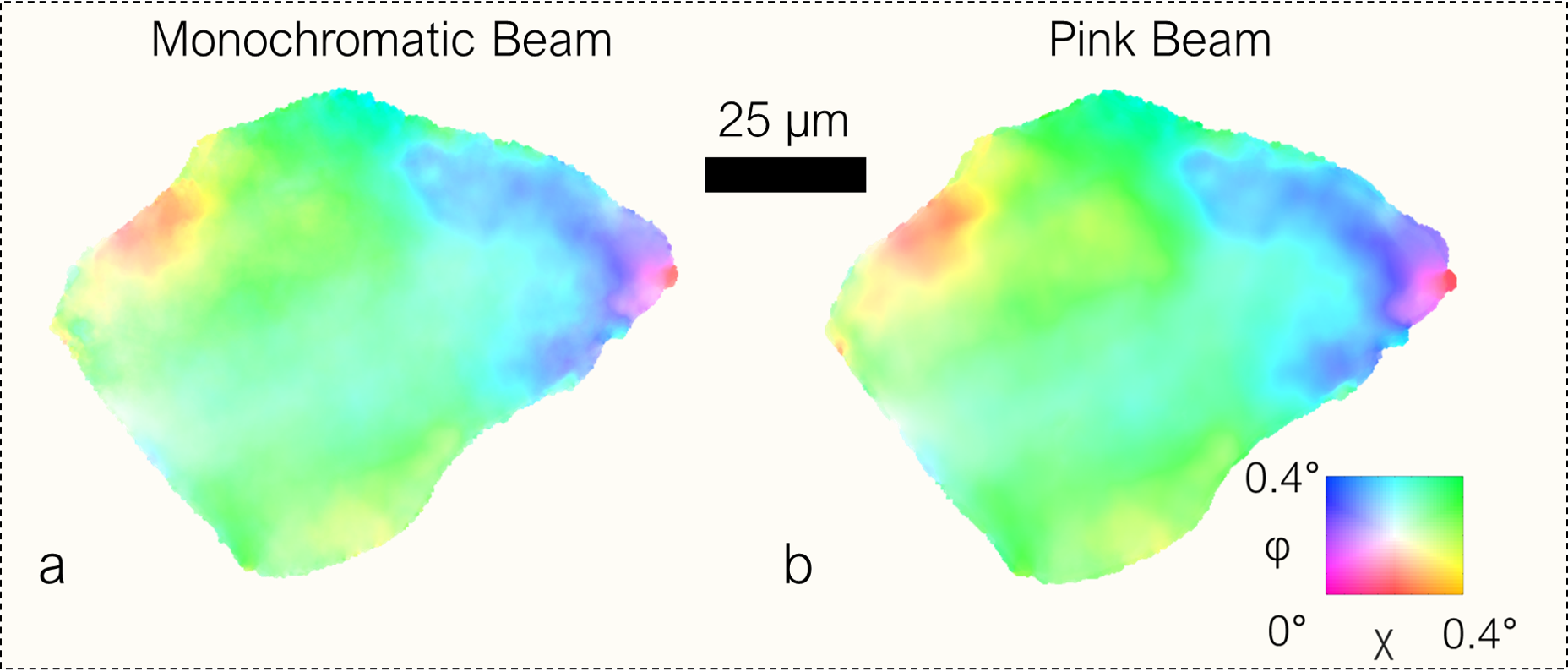} 
    \caption{\small \justifying \textbf{Comparison of orientation maps obtained with monochromatic and pink-beam DFXM.} 
    Projection Mosaicity maps representing local orientation variations, including orthogonal tilt components, for monochromatic (a) and pink-beam DFXM (b). 
    Mosaicity is defined by the two tilt components forming a 2D angular mesh around the \(111\) diffraction vector. 
    The color key indicates the angular range of local orientation variations. 
    }
    \label{fig:orientation_comparison}
\end{figure}
\subsection*{\textit{Intragranular Orientation Mapping Across Varying Defect Densities}}
\vspace{1.5mm}

\begin{figure*}[t]
    \centering
    \includegraphics[width=0.9\textwidth]{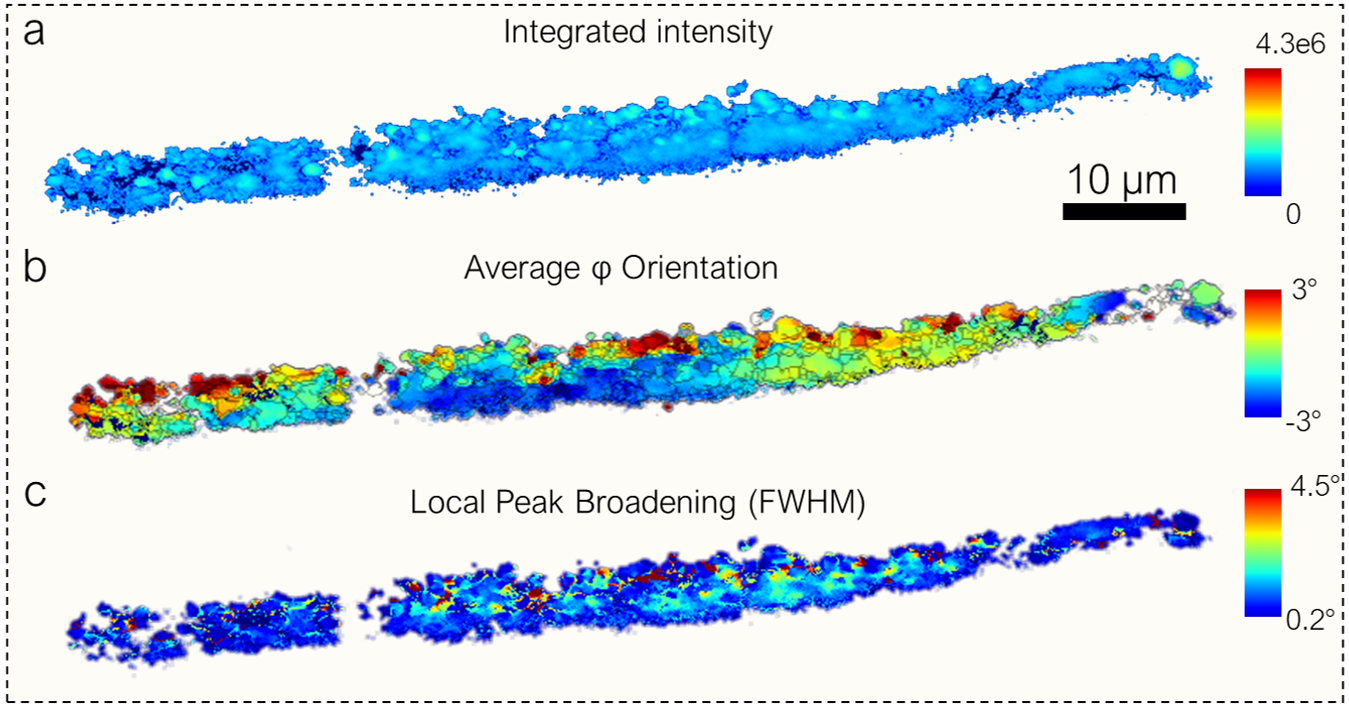} 
    \captionsetup{justification=justified, singlelinecheck=false, font=small}
    \caption{\textbf{Projection \pDFXM{} maps showing subgrain structure and orientation mapping in highly deformed ferritic Fe-3\%Si.} 
    Projection orientation maps were obtained around the (110) Bragg peak using dark-field X-ray microscopy (DFXM) through the objective lenses at a photon energy of 19 keV in a cold-rolled ferritic Fe-3\%Si sample with 50\% thickness reduction, indicative of high plastic deformation and elevated defect density. (a) Integrated intensity map, representing the amplitude (log scale) of Gaussian fits to the rocking curve~\cite{darfix}, i.e. tilting goniometer angle $\phi$ around the diffraction vector and collecting images at each point, highlights variations in diffracted signal strength. (b) The average $\phi$ orientation (peak position) map reveals local lattice rotations associated with deformation-induced substructure, with the cell structure overlaid using kernel average misorientation (KAM) filters to highlight cell boundaries. (c) Full-width at half-maximum (FWHM) of the rocking curve, i.e. local peak broadening, showing dislocation cell boundaries as regions with increased FWHM and reduced integrated intensity, indicative of strain localization and subgrain formation.}
    \label{fig:rolled_ferrite}
\end{figure*}

Next, we examine a recrystallized grain in a partially recrystallized Al1050 sample, imaged using both monochromatic and pink-beam DFXM. The goal is to assess how well \pDFXM{} preserves orientation information compared to conventional monochromatic DFXM while benefiting from significantly reduced acquisition times (by a factor of $\times$ 20>).  

Figure~\ref{fig:orientation_comparison} directly compares the two imaging modes, showing 2D orientation maps (\(\phi\) and \(\chi\) tilts over 0.4$\degree$ ranges) obtained with monochromatic DFXM (left) and \pDFXM{} (right). These maps represent a 2D angular grid at a fixed scattering angle, revealing local intragranular orientations around the (111) diffraction vector.  

Despite the broader energy bandwidth of \pDFXM{}, the orientation maps in Figures~\ref{fig:orientation_comparison}a and b exhibit nearly identical features to those obtained with monochromatic DFXM, demonstrating that \pDFXM{} preserves sufficient angular resolution to resolve intragranular misorientations. These findings confirm that \pDFXM{} provides reliable intragranular orientation mapping with minimal loss of angular resolution while significantly reducing acquisition times.
A notable observation is the increased mosaicity near the grain boundary (edges of the visualized grain), where interactions with neighboring grains induce local lattice distortions. These distortions appear as sharp orientation gradients at the periphery, which gradually diminish toward the grain interior. This suggests that even in a recrystallized state, the grain retains small but measurable orientation variations, likely reflecting its deformation and annealing history. During recrystallization, stored energy redistribution and grain boundary migration generally lead to structural homogenization \cite{humphreys_2004}. However, the persistence of residual misorientation within the grain indicates that local strain fields, possibly influenced by adjacent grains, continue to affect the microstructure. The enhanced mosaicity near the boundary suggests that grain growth dynamics may be influenced by these local distortions, which are well known to impact recrystallization kinetics in polycrystalline materials.

Let us now examine a contrasting case, one with a significantly higher defect density,by analyzing a highly deformed ferritic grain in a cold-rolled Fe-3\%Si sample that has undergone 50\% thickness reduction.

Figure~\ref{fig:rolled_ferrite} presents the subgrain structure and orientation mapping in this deformed ferritic Fe-3\%Si sample. The maps were obtained around the (110) Bragg peak using \pDFXM\ at a photon energy of 19~keV. Figure~\ref{fig:rolled_ferrite}a shows the integrated intensity of the diffracted signal, revealing local variations in diffraction contrast due to strain localization. Figure~\ref{fig:rolled_ferrite}b presents the peak position map, highlighting lattice rotations associated with deformation-induced substructure formation. A Kernel Average Misorientation (KAM) filter was applied to the peak position map and overlaid to highlight individual dislocation cell walls. More information about the KAM filtering is given in the SI. Finally, Figure~\ref{fig:rolled_ferrite}c displays the full-width at half-maximum (FWHM), local peak broadening of the rocking curve, which serves as an indicator of stored energy and dislocation cell boundary formation. 

Plastic deformation primarily occurs through the creation and motion of dislocations, which tend to arrange into cellular structures characterized by wall comprising high numbe rof dislocations and regions with lower dislocation density~\cite{mughrabi1986long,hansen2011deformed}. The development and evolution of these structures play a crucial role in determining the mechanical behavior of crystalline materials. Investigating the formation mechanisms of these organized dislocation networks remains a key focus of ongoing research, both through numerical simulations\cite{madec2003role} and experimental studies~\cite{zelenika20243d}. Based on our \pDFXM{} results, average $\phi$ orientation(i.e. center-of-mass) cell map with a KAM filter shown in Figure~\ref{fig:rolled_ferrite}b, the average cell size is estimated as 0.9~$\mu$m. This corresponds to an equivalent von Mises strain of $\epsilon_{vm} = 0.8$. While this estimate is based on a 2D projection and does not account for potential height variations in the 3D structure, the observed microstructural features in these examples are consistent with expectations from deformed grains. The corresponding equivalent von Mises strain of $\epsilon_{vm} = 0.8$ aligns with previously reported values for cold-rolled materials \cite{hansen2001microstructural}, though care should be taken in interpreting these values given the inherent limitations of 2D projections in capturing the full spatial arrangement of the microstructure.
The evolution of dislocation structures during cold rolling follows well-established mechanisms, wherein dislocations initially distribute homogeneously but gradually form well-defined subgrain boundaries as strain accumulates~\cite{hansen_scaling_prl, hansen_scaling_medium_high}. Dislocations arrange into low-energy configurations, leading to the development of cell walls and subgrain networks. As shown in Figure~\ref{fig:rolled_ferrite}, the FWHM map reveals regions of increased broadening, corresponding to dislocation cell walls where misorientation across boundaries increases with strain. This observation is consistent with established scaling relationships in deformation microstructures, where higher applied strains result in reduced cell sizes and greater misorientation across sub-boundaries~\cite{kestens_texture_formation}.
\begin{figure*}[t]
    \centering
   \includegraphics[width=1\textwidth]{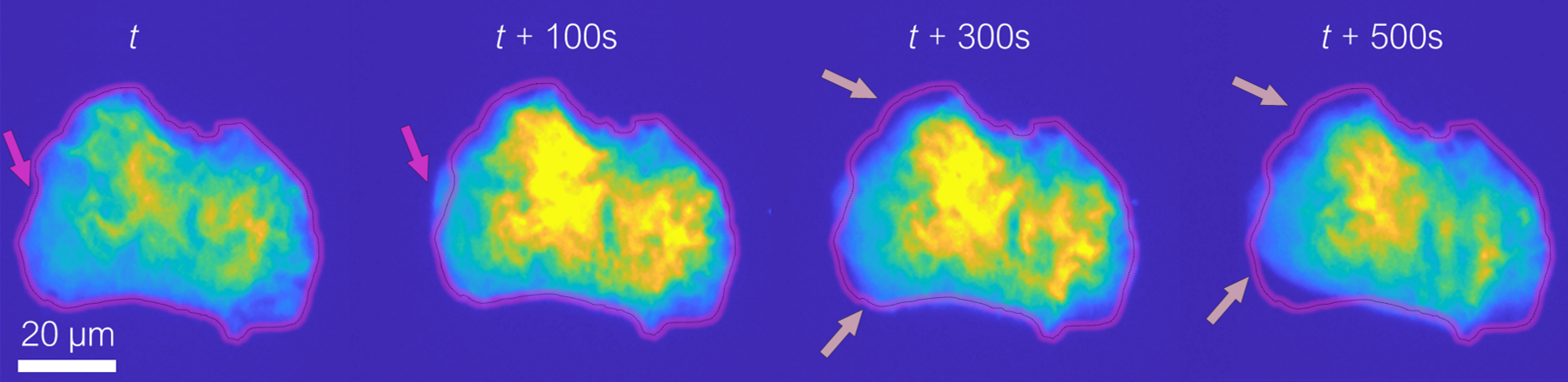} 
\caption{\small \textbf{In-situ observation of aluminum grain growth during isothermal annealing using single-frame projection imaging .}  
Time evolution of a single grain at \( t \), \( t+100 \)~s, \( t+300 \)~s, and \( t+500 \)~s.  
The false-color map represents diffracted intensity, where brighter regions correspond to higher intensity.   The pink dashed line marks the grain boundary at each time step, providing a reference for changes in grain shape and size.  
The outer boundary at \( t \) is overlaid on all subsequent time steps to highlight grain evolution relative to its initial state.  
Arrows indicate significant grain boundary motion: pink for growth and light pink for shrinkage.  These images are single DFXM projections, acquired at fixed \( \phi \) and \( \chi \) using a pink beam, with 400~ms exposure at 450°C.  
No angular scans were performed; the intensity reflects the integrated diffraction signal from the illuminated volume at the given orientation.  
The initial time step (\( t \)) corresponds to more than 1000 s of annealing, as detailed in the SI. A movie showing the full grain boundary evolution is provided in the supplementary videos. Further details on the annealing history and measurement conditions are given in the SI.  
}
     \label{fig:gg}
\end{figure*}

In body-centered cubic (BCC) metals such as ferritic iron, plasticity is governed by the operation of multiple slip systems, primarily along $\langle 111 \rangle$ directions on \{110\}, \{112\}, and \{123\} planes~\cite{kestens_texture_formation}. The complex interplay between these slip systems leads to the formation of various dislocation arrangements, including incidental dislocation boundaries (IDBs) and geometrically necessary boundaries (GNBs), which facilitate strain accommodation. The dislocation densities in these highly deformed microstructures can reach values on the order of $10^{15}$~m$^{-2}$, contributing to pronounced hardening and reduced ductility~\cite{hansen_scaling_prl}. The GND density in  the grain shown in Figure~\ref{fig:rolled_ferrite} is on the order of $10^{14}$~m$^{-2}$ calculated using the misorientation and average dislocation cell dimensions and spacing.

The ability of \pDFXM{} to resolve these dislocation structures is particularly valuable for studying deformation mechanism and related thermally activated phenomena such as recovery, recrystallization and grain growth. The projection image shown in Figure.~\ref{fig:rolled_ferrite} was collected using a 400~$\mu$m~$\times$~400~$\mu$m box beam, allowing comprehensive mapping of the deformed structure. Performing the same measurements with a monochromatic beam did not yield sufficient intensity to capture the diffracted signal from these heavily deformed grains. 
A comparison of the diffracted signal from this grain in the near-field, obtained using both a monochromatic beam and a pink beam, is presented in the SI. These observations highlight the advantages of \pDFXM{} for investigating deformation-induced substructures in crystals. Its ability to probe high-strain regions with exceptional spatial and angular resolution enables detailed studies of work hardening, subgrain evolution, and annealing phenomena in plastically deformed materials. Moreover, hierarchical structures, such as biominerals and geological samples from earth and different planets, often exhibit textured features with broadened diffraction peaks, making them promising candidates for study using \pDFXM{}'s increased flux.

It is important to note that at photon flux levels approaching $10^{15}$~photons/sec, thermal effects become a concern for high-$Z$ materials such as iron, particularly when sample dimensions are insufficient to dissipate heat via conduction. To mitigate beam heating, an airflow system was implemented to maintain thermal stability during data acquisition. This consideration is crucial for \textit{operando} studies, where prolonged exposure to high-intensity beams may introduce artifacts due to thermal expansion, recovery or even recrystallization.

     \begin{figure*}[t]
    \centering
    \includegraphics[width=2\columnwidth]{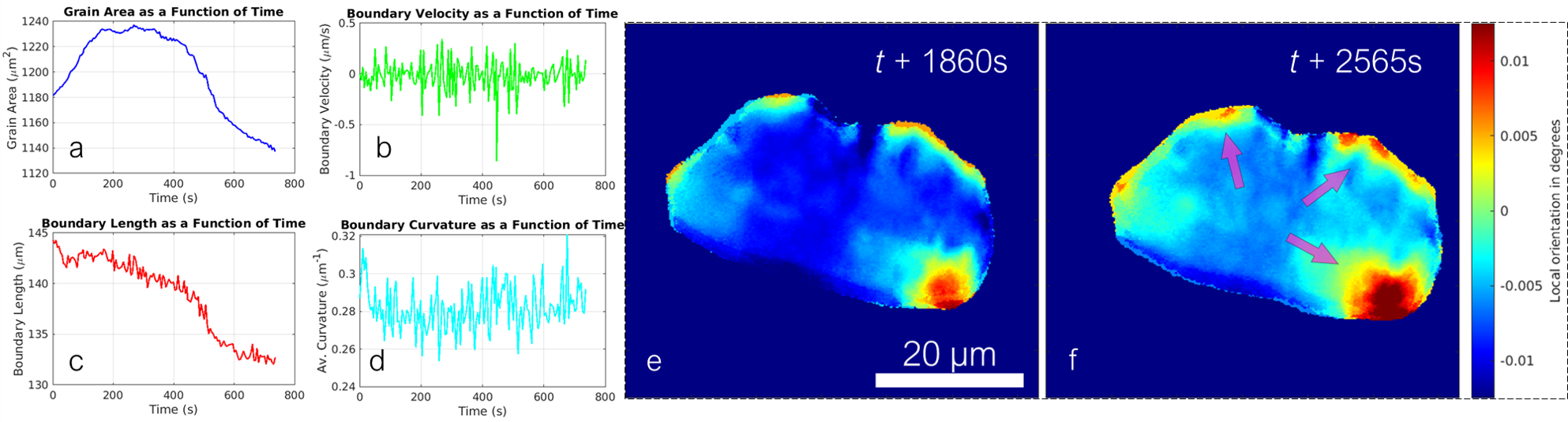} 
    \caption{\small \justifying \textbf{Evolution of Grain Boundary Statistics and Internal Misorientation During Isothermal Annealing via Live Orientation Mapping} 
    (a) Grain area as a function of time, showing an initial increase followed by a decrease. 
    (b) Boundary length evolution, which does not directly correlate with grain area, indicating irregular boundary dynamics. 
    (c) Boundary velocity fluctuations, with noticeable spikes at certain times, suggesting a stop-and-go motion of the grain boundary, consistent with observations reported elsewhere\cite{jensen2020}
    (d) Average boundary curvature as a function of time, showing variations but no direct correlation with grain area, suggesting that curvature is influenced by local irregularities rather than simple growth kinetics. 
    The boundary overlay used for these measurements was calculated using a thresholding method, with further details provided in the SI. Panels (e) and (f) show the reconstructed average $\phi$ orientation map of the (111) reflection in projection mode, obtained using box-beam illumination, providing a 2D projection of the 3D grain.  
Between \( t+1860 \)~s and \( t+2565 \)~s, the development of local misorientation is observed, attributed to mechanical forces exerted by neighboring grains during recrystallization and growth.
}
    \label{fig:grain_stats}
\end{figure*}

\subsection*{\textit{Grain Growth Dynamics during Isothermal Annealing}}
\vspace{1.5mm}
To demonstrate the in situ monitoring capabilities of \pDFXM{}, we track grain evolution during recrystallization annealing of cold-forged aluminum, enabling direct observation of grain boundary migration and structural changes over time. The ability to capture real-time changes in grain morphology and orientation provides a unique opportunity to quantify microstructural evolution at an unprecedented level of detail. This experiment serves as a benchmark for the level of quantitative analysis now becoming available with \pDFXM{}, offering insight into dynamic recrystallization processes. The analysis focuses on a single grain, mapping its boundary motion and deformation-driven heterogeneities. The sample was first annealed to $\approx$ 29.3\% global recrystallization (see SI for details), after which a grain of interest was selected and its evolution tracked using \pDFXM{} during further annealing.
\par
In Figure~\ref{fig:gg} we observe the microstructure of the GOI at different annealing times in single frames during isothermal annealing. Brighter regions in the 2D \pDFXM\ images indicate higher diffracted intensity. These images were acquired using box-beam illumination at a fixed sample tilt, with a continuous scanning rate of 400 ms per frame. The experiment was conducted at a synchrotron ring current of 35 mA, seven times lower than the standard 200 mA mode typically used for monochromatic DFXM. Under standard conditions, much higher frame rates would be possible, reducing scan times from 400 ms to the tens of milliseconds range, significantly enhancing temporal resolution. Despite the lower flux, real-time tracking of grain boundary evolution was achieved.

The grain boundary position, marked by a pink dashed line, enables direct comparisons of shape and size changes across different time steps. The outer boundary at time $t$ is overlaid on all subsequent images as a reference for quantifying grain evolution. Arrows indicate boundary motion, with pink arrows highlighting growth regions and light pink arrows marking shrinkage zones. Snapshots at different annealing times, capture the dynamic nature of boundary migration. The changes in grain size and morphology suggest that grain growth is not uniform but occurs through localized movements, potentially influenced by local strain fields, effect of neighboring grains, and impurity distributions. A corresponding video showing the full temporal evolution of the grain during annealing is provided in the Supplementary Videos.

We now turn to the analysis of grain boundary dynamics using statistical measures extracted from the time-resolved mapping of the recrystallized grain. Figure~\ref{fig:grain_stats} presents the evolution of grain boundary properties during isothermal annealing, based on the same grain analyzed in Figure~\ref{fig:gg}. The data provides insight into the complex nature of boundary motion, revealing deviations from conventional grain growth models. 

Figure~\ref{fig:grain_stats}a shows the evolution of the grain area over time. Initially, the area increases, reaches a maximum, and then decreases, suggesting that the grain expands before being consumed by its neighbors. However, as seen in Figure~\ref{fig:grain_stats}b, the boundary length does not exhibit a direct correlation with area, indicating that the grain boundary undergoes local irregular deformations rather than uniform shrinkage.

Fig.~\ref{fig:grain_stats}c plots the boundary velocity as a function of time, revealing notable fluctuations with clear spikes. These spikes indicate intermittent boundary movement, often referred to as stop-and-go motion, a behavior previously observed in experimental studies by eg. Zhang and Jensen \cite{jensen2020}. This suggests that the boundary does not migrate at a steady rate but instead experiences moments of acceleration and stagnation, likely due to interactions with local microstructural heterogeneities or impurity drag effects.  Figure~\ref{fig:grain_stats}d shows the evolution of the average curvature of the grain boundary over time. While fluctuations are evident, no clear correlation with grain area is observed, suggesting that boundary migration is primarily influenced by local irregularities rather than a simple size-dependent, curvature-driven mechanism. A similar observation was reported in other polycrystals, where 3DXRD measurements demonstrated that grain boundary curvature does not directly correlate with boundary velocity during growth~\cite{zhang2020grain}. Based on the fitted Avrami curves, recrystallization had already almost reached 100\% completion by the time our live scans began. The annealing profile and Avrami curves (provided in the SI) indicate that our first scans started around 1000 seconds and continued until nearly 2000 seconds. As a result, Figures~\ref{fig:grain_stats}5 and~\ref{fig:grain_stats}6a-d primarily capture the final stages of recrystallization or the transition into grain growth, while Figures~\ref{fig:grain_stats}6e-f explicitly show microstructural evolution during grain growth, after recrystallization was fully completed.

It has been previously reported that nuclei with new orientations relative to the deformed state can form during recrystallization, that grain boundaries do not move at a constant rate but rather in a stop-and-go fashion, even in weakly deformed single crystals, and that the surfaces of recrystallizing grains exhibit roughness with characteristic protrusions \cite{jensen2020}. Additionally, each recrystallizing grain was found to have its own kinetics and activation energy. These findings highlight the complexity of the mechanisms behind the growth of recrystallizing grains  beyond classical smooth-boundary models \cite{lauridsen2000kinetics,poulsen2011situ}.

Our results for this particular grain align with these observations. The fluctuations in boundary velocity confirm the non-uniform motion of the grain boundary, showing distinct stop-and-go behavior. Similarly, the observed variations in curvature further support the idea that local irregularities play a role in the migration process. The lack of direct correlation between boundary length and grain area suggests that deformation-driven inhomogeneities in the microstructure may influence boundary movement, reinforcing previous experimental observations.  Additionally, the presence of particles in the alloy may contribute to this stop-and-go behavior by pinning the boundary migration.

A key finding of this study is the observation of local orientation buildup within a recrystallized grain, even after extended annealing times, as shown in Figure~\ref{fig:grain_stats}e and f. These high-resolution orientation maps, reconstructed from rapid rocking curve acquisitions, reveal that misorientations initially form at grain boundaries and progressively extend into the grain interior. The fact that this buildup persists after 12,000 seconds of annealing demonstrates that recrystallized grains do not immediately reach a uniform, strain-free state but continue evolving under the influence of neighboring grains.

Looking at Figure~\ref{fig:grain_stats}e, at $t$ + 1860s, the orientation distribution appears more homogeneous, but fluctuations remain prominent along the grain boundaries, particularly on the lower right side of the grain. Over the next 700 seconds, increasing mechanical interactions with neighboring grains lead to greater lattice distortion. Initially concentrated at the grain boundaries, this effect gradually propagates into the grain interior, resulting in a non-zero residual distortion. This is particularly relevant for materials science, as residual strain in recrystallized grains is a critical factor in mechanical performance and is often present in real applications~\cite{zhang2022local}.

The conventional understanding of recrystallization assumes that newly formed grains rapidly reach equilibrium and become strain-free\cite{humphreys_2004}. However, recent studies challenge this assumption, showing that recrystallized grains can retain residual strain and continue evolving during annealing. For instance, research on partially recrystallized aluminum has demonstrated significant local residual strain variations within individual grains, indicating that lattice distortions persist beyond initial grain growth~\cite{lindkvist20233d}. These findings align with our observations on Figure \ref{fig:grain_stats}e-f, and suggest that lattice distortions (i.e. observed local orientation buildup) and potentially residual strain are not fully eliminated during recrystallization but are influenced by interactions with neighboring grains and local stress fields.

Movies showing the local orientation and FWHM buildup during this annealing step are provided in the Supplementary Videos.

Our in situ recrystallization and grain-growth study using \pDFXM{} highlights the complexity of grain boundary migration, emphasizing the role of local variations in boundary shape, velocity fluctuations, and stop-and-go motion. More broadly, \pDFXM{} enables real-time, high-resolution tracking of microstructural evolution in bulk materials, providing direct experimental access to recrystallization kinetics and defect interactions that remain challenging to resolve with conventional methods.

\section*{Conclusions and Outlook}

In this work, we have demonstrated Pink-Beam Dark Field X-ray Microscopy (\pDFXM{}) at the ID03 beamline of the European Synchrotron Radiation Facility (ESRF), achieving 100 nm spatial resolution with a 27-fold increase in diffracted intensity compared to monochromatic DFXM. This enables significantly faster acquisition times and high-resolution imaging, making \pDFXM{} a viable method for real-time, high-throughput studies of deformation microstructures, grain growth, and phase transformations in bulk crystalline materials.

A direct comparison with monochromatic DFXM confirms that \pDFXM{} retains sufficient angular resolution to resolve intragranular structures, providing non-destructive, high-resolution 3D mapping of recrystallized grains in industrially relevant samples. The ability to switch rapidly between monochromatic and pink-beam modes provides experimental flexibility, allowing high-intensity imaging with \pDFXM{} while preserving the option to map elastic strains using a monochromatic beam when needed.

Beyond this first demonstration, \pDFXM{} extends the capabilities of diffraction-contrast imaging to previously inaccessible materials and processes. Future applications and key considerations include:
\begin{itemize}
    \item \textit{Operando Studies Across Multiple Domains:} \pDFXM\ enables real-time observation of microstructural evolution in catalytic processes, battery degradation, phase transformations, fatigue, hydrogen charging, and beyond. Its high temporal resolution makes it uniquely suited for studying dynamic changes in energy materials and structural alloys under real-world conditions.

    \item \textit{Expanding the Study of Challenging Materials:} For the first time, \pDFXM\ enables 3D imaging of highly textured crystals, poorly diffracting semi-crystalline materials, and hierarchically ordered crystalline systems with 100 nm spatial resolution and a large field of view (~100 µm). This capability extends diffraction-contrast imaging to materials with poor diffraction signal such as biominerals, highly deformed metals, and semi-crystalline polymers.

    \item \textit{Correlative Imaging with Complementary Techniques:} The integration of \pDFXM\ with texture tomography, scanning 3DXRD, and DCT will provide multi-scale insights into polycrystalline materials, while rapid acquisition of grains of interest will facilitate in-depth dynamic studies.

    \item \textit{Beam Damage Mitigation: }While beam damage remains a consideration, particularly for thin samples, it can be mitigated using higher-energy X-rays to reduce absorption\cite{lawrence2021}. The beamline’s capability of operating up to 60 keV, along with upcoming high-energy imaging optics such as diamond compound refractive lenses, will further enhance the study of thicker and denser materials. Additional strategies such as cryostream cooling or air/N\textsubscript{2} flow can be employed to minimize beam heating effects.

    \item \textit{Advancements in X-ray Optics:} The development of diamond lenses with correction plates will be critical for improving high-energy \pDFXM. Diamond’s resistance to pink-beam exposure, combined with correction plates that reduce lens aberrations, will significantly enhance image quality and broaden the range of materials that can be studied.

    \item \textit{Theoretical and Computational Advancements}: A more rigorous framework for understanding bandwidth effects is needed, including theoretical comparisons with geometric formalisms of ideal and distorted crystals. Additionally, machine learning algorithms will be applied for real-time image optimization, using monochromatic images and simulations as training datasets. This work is currently in progress and will further expand \pDFXM’s capabilities.

\end{itemize}

\section*{Methods}
\subsection*{Samples}

Multiple sample types were used in this study, each selected to match specific case studies investigating different microstructural conditions. 
The AA1050 aluminum samples, with a 1 $\times$ 1 mm$^2$ cross-section and a length of a few millimeters, were prepared using Electrical Discharge Machining (EDM) to preserve the microstructure. The partially recrystallized samples were annealed \textit{ex situ} at \SI{325}{\degreeCelsius} to study the evolution of the microstructure. One of these samples was then cold-forged using unidirectional forging, reducing its thickness by 50\%, for subsequent \textit{in-situ} annealing study.

The cold-rolled ferritic Fe-Si alloy was used to examine highly deformed grains, while other samples were utilized for analyzing recrystallized and partially recovered microstructures. The material was first hot rolled to 2.0~mm, annealed, and then subjected to a 50\% thickness reduction through cold rolling, reaching a final thickness of 1.0~mm. To minimize surface effects, the as-cold-rolled sample was manually ground to remove the top 15\% of its thickness before being machined into small stick-shaped specimens. These specimens, measuring \(7.0 \times 0.150 \times 0.200\)~mm\(^3\) (rolling direction (RD) \(\times\) transverse direction (TD) \(\times\) normal direction (ND)), were extracted parallel to the rolling direction using spark erosion wire cutting to minimize mechanical damage. This sample was specifically chosen for investigating the evolution of dislocation cell structures and deformation-induced subgrain formation.

\subsection*{DFXM}
Our DFXM experiments were carried out at Beamline ID03 at the European Synchrotron Radiation Facility (ESRF) \cite{isern2024}. We used two selected photon energies, 17 and 19\un{keV}. These wavelengths were picked by the double multilayer monochromator (DMM) and the channel cut crystal monochromator (CCM) for the pink and monochromatic beam, respectively. We used a parallel beam, shaped by the slits on the order of hundreds of micrometers depending on the grain size.The photon flux incident on the sample after the beam-defining slits was measured using an ion chamber filled with N$_2$ gas. The estimated flux values are:

\begin{itemize}
    \item $8.1 \times 10^{11}$ photons/s for the monochromatic beam,
    \item $3.71 \times 10^{12}$ photons/s for the transfocator-focused beam,
    \item $1.02 \times 10^{14}$ photons/s for the pink beam.
\end{itemize}

These values correspond to Fig.~\ref{fig:int_comp} (a), (b), and (c), respectively. 

It is important to note that these measurements were performed at a ring current of 35 mA with a slit size of 0.5 × 0.5 mm$^2$. Under standard experimental conditions, the ring current is typically around 200 mA.

A near-field alignment camera was positioned 40~mm behind the sample to orient the crystal to the Bragg condition. Once alignment was achieved, the camera was removed, and imaging was performed using an objective X-ray lens consisting of 87 bi-paraboloid Be lenslets (2D focusing optics), each with a radius of curvature of $R=50~\mu$m. The compound refractive lens (CRL) entry plane was placed between 260 and 320 mm downstream of the sample along the diffracted beam path depending on the energy. The numerical aperture is calculated to be 0.693 mrad at 17 keV with the X-ray magnification of 18.35 using the Equation 9 in the ref\cite{Poulsen2017}. The CRLs were aligned using an ionchamber at the exit of the lens box.

The objective lens magnified the diffracted image of the sample, projecting it onto the far-field detector with an X-ray magnification of $M_{\text{x}} = 14.38$-–$18.1\times$, depending on the energy. This setup enabled the projection imaging of diffracting grains throughout their entire volumes, capturing their structural evolution in the far field.

Our far-field imaging detector used an indirect X-ray detection scheme. This detector consisted of a scintillator crystal, a visible light microscope, and a PCO.edge 4.2 bi (back-illuminated) sCMOS camera with $2048 \times 2048$ pixels. It was placed 5010\un{mm} from the sample. The visible light optics inside the far-field detector could switch between $10\times$ and $2\times$ magnification to achieve an effective pixel size of 0.65\un{\mu m} or 3.25\un{\mu m} at the scintillator, and 36.3\un{nm} or 182\un{nm} at the sample position, respectively. 

Two types of scans were performed in this work: rocking scans, mosaicity scans . The rocking scans acquired images while scanning the tilt angle $\phi$, see Fig.~\ref{fig:dfxm_schematics}.
Mosaicity scans were collected to more thoroughly by measuring distortions along the two orthogonal tilts $\chi$ and $\phi$, cf. Fig.~\ref{fig:dfxm_schematics}. 
With this data, each voxel can be associated with a subset of a $(111)$ pole figure of Aluminum, allowing us to generate Center of Mass (COM) maps to describe the average direction of the $(111)$ orientation for each voxel in the layer \cite{darfix}. We note that the angular resolution in the COM maps is substantially better than the step size.

\subsection*{In-Situ Annealing}

The sample underwent two interrupted annealing steps. During the first step, the sample was heated for 38 seconds while live diffraction images were collected. After this initial heating, the sample was rocked over a 4-degree range. The recrystallization percentage was estimated to be approximately 4\%, based on the intensity of the diffraction spots.

A second heating step followed, bringing the total annealing time to 105 seconds. Afterward, the X-ray energy was adjusted from 33 keV to 17 keV. A recrystallized grain of interest was identified and monitored in the far field using pink beam illumination through the objective. More information about the full annealing temperature time profile is given in SI.

\subsection*{Data Analysis Methods}

We include the MATLAB and Python scripts and functions used in this work in the Github folder available at \href{https://github.com/cyil-esrf/D-REX}{https://github.com/cyil-esrf/D-REX}.  
Additional data processing was carried out using the \textit{darfix} package \cite{Garriga2023}.

\section*{Acknowledgement}
We thank ESRF for providing the beamtime at ID03. CY, AS, LL and VS acknowledge the financial support from the ERC Starting Grant nr 10116911. HFP acknowledges support from the ERC Advanced Grant nr 885022 and from the Danish ESS lighthouse on hard materials in 3D, SOLID.  C.Y. acknowledges the technical help provided by H. Isern and T. Dufrane during the experiments. 

\section*{Author contributions statement}

C.Y., C.D., and H.F.P. designed the study and planned the experiments. N.M. and Y.Z. prepared the samples. C.Y., M.S., Y.L., V.S., and C.D. carried out the experiments. C.Y., A.S., and L.L. analyzed the data. C.Y. drafted the manuscript with contributions from A.S. and L.L. All authors discussed the results and reviewed the manuscript.

\section*{Data Availability}
The datasets used and/or analyzed during the current study are available from the corresponding author on reasonable request.
\bibliography{references}

\end{document}